\documentclass[twocolumn,showpacs,preprintnumbers,amsmath,amssymb,superscriptaddress]{revtex4-1}

\usepackage{graphicx}
\usepackage{dcolumn}
\usepackage{bm}
\usepackage{dsfont}
\usepackage{color,mathptm,times}

\DeclareMathOperator\erf{erf}

\begin{document}

\preprint{}

\title{Violation of Bell inequalities in larger Hilbert spaces: robustness and challenges}

\author{Werner Weiss}
\email{werner.weiss@uni-ulm.de}
\affiliation{Institute for Complex Quantum Systems \& Center for Integrated Quantum Science and Technology, Ulm University, Albert-Einstein-Allee 11, 89069 Ulm, Germany}
\author{Giuliano Benenti}
\affiliation{Center for Nonlinear and Complex Systems,
Universit\`a degli Studi dell'Insubria, via Valleggio 11, 22100 Como, Italy}
\affiliation{Istituto Nazionale di Fisica Nucleare, Sezione di Milano,
via Celoria 16, 20133 Milano, Italy}
\author{Giulio Casati}
\affiliation{Center for Nonlinear and Complex Systems,
Universit\`a degli Studi dell'Insubria, via Valleggio 11, 22100 Como, Italy}
\affiliation{International Institute of Physics, Federal University of
Rio Grande do Norte, Natal, Brasil}
\author{Italo Guarneri}
\affiliation{Center for Nonlinear and Complex Systems,
Universit\`a degli Studi dell'Insubria, via Valleggio 11, 22100 Como, Italy}
\affiliation{Istituto Nazionale di Fisica Nucleare, Sezione di Pavia, Via Bassi 6, 27100 Pavia, Italy}
\author{Tommaso Calarco}
\affiliation{Institute for Complex Quantum Systems \& Center for Integrated Quantum Science and Technology, Ulm University, Albert-Einstein-Allee 11, 89069 Ulm, Germany}
\author{Mauro Paternostro}
\affiliation{Centre for Theoretical Atomic, Molecular, and Optical Physics, School of Mathematics and Physics, Queen's University, Belfast BT7 1NN, United Kingdom}
\author{Simone Montangero}
\affiliation{Institute for Complex Quantum Systems \& Center for Integrated Quantum Science and Technology, Ulm University, Albert-Einstein-Allee 11, 89069 Ulm, Germany}


\date{\today}

\begin{abstract}
We explore the challenges posed by the violation of Bell-like inequalities by $d$-dimensional systems exposed to
imperfect state-preparation and measurement settings. 
We address, in particular, the limit of high-dimensional systems, naturally arising when exploring the quantum-to-classical transition.  
We show that, although suitable Bell inequalities can be violated, in principle, for any dimension of given subsystems, it is in practice increasingly challenging to detect such violations, even if the system is prepared in a maximally entangled state. We characterize the effects of random perturbations on the state or on the measurement settings, also quantifying the efforts needed to certify the possible violations in case of complete ignorance on the system state at hand.
\end{abstract}

\maketitle

\section{Introduction} 

Year 2014 has marked the 50$^{\rm th}$ anniversary of John S. Bell's paper~\cite{bell_einstein_podolsky_rosen_1964} on the Einstein-Podolski-Rosen paradox~\cite{einstein_can_1935}, a landmark achievement that has shaped the research in modern quantum mechanics. Since then, Bell inequalities, which have proven the incompatibility of quantum mechanics with any local realistic theory, have been established as a viable tool for deepening our understanding of the nature of reality and, from a more pragmatic perspective, identify unambiguously correlations of a non-classical nature~\cite{brunner2014}. 

Non-locality tests for discrete-variable systems have been formulated, over the years, based on both dichotomic and multichotomic local measurement settings~\cite{kaszlikowski}, showing that bipartite $d$-level entangled states are in conflict with local realistic assumptions more strongly than entangled qubits. Inequalities for bipartite $d$-dimensional systems have been put forward~\cite{collins_bell_2002}, while various formulation for multipartite $d=2$ systems have been proposed~\cite{multi}, and extended to generic arbitrary dimensional systems~\cite{Son}. Continuous-variable Bell-like inequalities have been derived for bipartite states using the phase-space formalism~\cite{banaszek}, and extended to multipartite scenarios recently~\cite{Lee}.

The investigation on non-locality of discrete large-$d$ systems is particularly relevant in light of the goal of inferring non-classical effects in increasingly large systems. The possibility to unambiguously signal non-classical correlations, strong enough to violate a Bell-like inequality, would provide undoubtable evidence of the relevance of quantum theory in the large-scale domain, and bring new perspectives to the investigation of the quantum-to-classical transition~\cite{zurek}. First steps along similar lines of investigation have been made by addressing the class of states dubbed 
as {micro-macro} ones~\cite{micromacro}, which have then been extended to macro-macro scenarios~\cite{macromacro}. 

When tackling the phenomenology of non-classical correlations in large-scale systems, the detrimental effects of improper measurements should be considered on the same footing as non-ideal state preparation. It has been shown that coarse-grained measurements do indeed affect the chances of falsifying local realistic models, even when dealing with otherwise ideal state resources~\cite{ralph}. On the other hand, a typical assessment of the robustness of Bell-like tests is based on the amount of white noise that a given resource state is able to tolerate before losing any discrepancy with local hidden variable (LHV) models~\cite{collins_bell_2002}. This approach was the basis for the conclusion drawn by Kaszilowski {\it et al.} on bipartite $d$-level systems~\cite{kaszlikowski}. However, while such analyses provide useful information on the effects that imperfections and environmental mechanisms have on the revelation of non-classicality, a systematic study of the scaling behavior of non-idealities against the {
\it size} of a given system has only been partially addressed~\cite{lask_2014,almeida_2007,Atkin_2015}. Yet, this is key information for the quest of observing non-classicality at the large-scale limit. For instance, as we address in this paper, Bell inequalities formulated for large-$d$ systems require a large number of measurement settings: their inadequate arrangement could have rather significant implications for the successful falsification of LHV models even when states endowed with a large degree of quantum correlations are considered. In this sense, our study goes well beyond currently available literature~\cite{kaszlikowski}.

Our study here addresses precisely this important issue, which we tackle in a careful quantitative manner. By addressing experimentally relevant imperfections in the Bell test of $d$-dimensional systems, we point out, through a careful statistical analysis, the existence of thresholds in the amount of randomness that a given resource state (or, equivalently, measurement setting) can tolerate. As we find out, such threshold, which scales unfavorably with the dimensions of the local systems being used, is rather severe and poses significant questions on the suitability of Bell-like tests in systems verging to the mesoscopic or macroscopic scale, when evaluated against the actual value taken by $d$.
In some conditions, even a maximally entangled two-qutrit state could fail to violate a Bell inequality built from slightly improper measurement settings.  

The remainder of this paper is organized as follows. In Sec.~\ref{sec:GenBell} we introduce the inequalities addressed in this work, setting the notation that will be used throughout our analysis. Sec.~\ref{sec:Stability} is devoted to the analysis of the robustness of such Bell-like tests for systems of increasing dimensionality and exposed to random perturbations of the otherwise ideal measurement settings required by each of the inequalities that we study. In Sec.~\ref{sec:Optimization} we extend our study to the case of random initial  states, showing the inherent difficulties related to running Bell tests on such resources, and discussing the effectiveness of more sophisticated statistical sampling techniques for ascertaining non-locality in randomly picked states of $d$-dimensional systems. Finally, in Sec.~\ref{conc} we draw our conclusions. 

\section{\label{sec:GenBell} Generalized Bell inequalities}

In this Section we review the formulation of Bell inequalities for $d$-dimensional systems
that are used hereafter. The original formulation of the Bell inequalities 
relates the joint probabilities for local measurements with two possible outcomes ($d=2$),
typically the polarization of photons or a spin-1/2 system (dimension $d$ and the spin quantum number $l$ obey the relation $d=2l+1$).
However, it is possible to extend (in a non unique way) such formulation to systems with higher local dimension~\cite{collins_bell_2002}:
We consider a coupled bipartite system and two parties (Alice and Bob) each of them having access to a single partition, of dimension $d$, of the system
(either subsystem $\mathcal{A}$ or $\mathcal{B}$) and able to measure two independent observables, $A_1$, $A_2$ and $B_1$, $B_2$, respectively.
The operators $A$s and $B$s are assumed to have the same spectrum, $A_i |A_i^\ell\rangle = \ell  |A_i^\ell\rangle$ and $B_j |B_j^\ell\rangle = \ell  |B_j^\ell\rangle$, with $\ell =0, \dots, d-1$.
Clearly, for $d$-dimensional subsystems the measurement can have $d$ different outcomes. The generalized Bell inequalities 
are obtained summing the probabilities that Alice and Bob obtain equal (or correlated) outcomes of their measurements. 
In particular, the first Bell expression we will consider reads as 
\begin{eqnarray}
 I \equiv& P(A_1 = B_1) + P(B_1 = A_2 + 1) 
\\ \nonumber
 & + P(A_2 = B_2) + P(B_2 = A_1),
 \label{Ibell}
\end{eqnarray}
where  $P(A_i = B_j)$ are the probability for Alice and Bob to have the same measure outcomes:
\begin{equation}
 P(A_i = B_j) = \sum\limits_{\ell=0}^{d-1} P(A_i= \ell, B_j = \ell),
\end{equation}
and similarly 
\begin{equation}
 P(A_i = B_j + k) = \sum\limits_{\ell=0}^{d-1} P(A_i = \ell, B_j = \ell + k\ \mathrm{mod}\ d).
\end{equation}
In general, one can show that for local theories only three of the four equalities in (\ref{Ibell}) can be satisfied at the same time, that is for local theories $I \le 3$, 
while non-local theories could reach a value of $I = 4$. 

Beside the generalization of the standard Bell inequality presented above, we also consider a 
second inequality, which at variance with the previous one is built with a number of terms that grows with the local dimension of the subsystems. Explicitly
\begin{eqnarray}
 I_d \equiv&& \sum\limits_{k=0}^{[d/2] - 1} \left( 1 - \frac{2k}{d-1} \right)  \nonumber \\ 
 &&\times \biggl\{ \Bigl[ P(A_1 = B_1 + k) + P(B_1 = A_2 + k + 1) \nonumber  \\
 &&+ P(A_2 = B_2 + k) + P(B_2 = A_1 + k) \Bigr] \nonumber \\
 &&- \Bigl[ P(A_1 = B_1 -k -1) + P(B_1 = A_2 -k) \nonumber \\
 &&+ P(A_2 = B_2-k-1) + P(B_2 = A_1 -k-1)  \Bigr] \biggr\} . 
\label{Idbell}
\end{eqnarray}
As before, one can show that the maximal possible value achievable by this quantity is $I_d = 4$. However, LHV models allow only for $I_d \le 2$.

\begin{figure}
\includegraphics[width=0.44\textwidth]{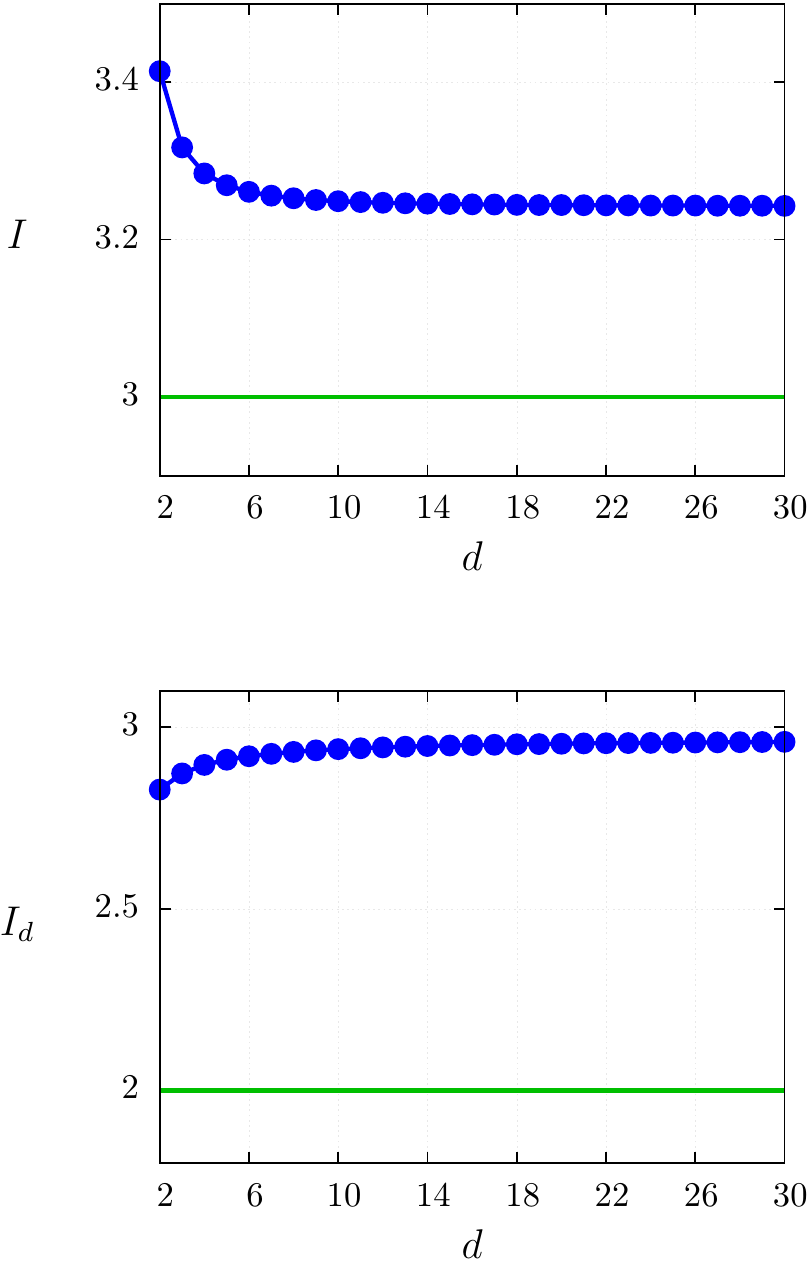}
\caption{(Color online) \label{fig1} Values for the generalized Bell expressions $I$ and $I_d$ plotted against the spin quantum number $l = (d-1)/2$ of each subsystem. In order to evaluate $I$ and $I_d$ we have used the state $|\Phi^+\rangle$ and the measurement operators given in Eqs.~(\ref{eq:optAxes}).
Bottom (green) lines: Maximum values achievable through local hidden variable models.}
\end{figure}

Generalized maximally entangled (Bell) quantum states 
of two $d$-dimensional systems,
\begin{equation}
 |\Phi^+\rangle = \frac{1}{\sqrt{d}}\sum\limits_{j=0}^{d-1} |j\rangle_{A} \otimes |j\rangle_{B},
 \label{bstate}
\end{equation}
violate these inequalities for any value of $d$~\cite{collins_bell_2002}. 
It has been shown~\cite{collins_bell_2002} that, for $|\Phi^+\rangle$, the observables that are suited to test the generalized Bell inequalities have
associated eigenstates of the form
\begin{subequations}
\label{eq:optAxes}
\begin{eqnarray}
|k\rangle_{A,a} =& \frac{1}{\sqrt{d}} \sum\limits_{j=0}^{d-1} \exp \left[ \mathrm{i} \frac{2\pi}{d} j (k + \alpha_a) \right] |j\rangle_A, \label{eq:optAxes1}
\\
|l\rangle_{B,b} =& \frac{1}{\sqrt{d}} \sum\limits_{j=0}^{d-1} \exp \left[ \mathrm{i} \frac{2\pi}{d} j (-l + \beta_b) \right] |j\rangle_B\label{eq:optAxes2}.
\end{eqnarray}
\end{subequations}
Here, $a,b=1, 2$ with $\alpha_1 = 0$, $\alpha_2 = 1/2$, $\beta_1 =-\beta_2 = 1/4$~\cite{collins_bell_2002}. Projections of the state of the system onto such states always leads to a violation of $I$ and $I_d$, independently of the particular value of $d$. This is shown in Fig.~\ref{fig1}, where we report the exact numerical values for both $I$ and $I_d$ as a function of the spin quantum number $l$ of the system. Note that, if the probabilities are computed according to the rules of 
quantum mechanics, the maximal value of $I$ achievable for $d=2$ is 
$I_{\rm max} = 2 + \sqrt{2} \approx 3.41$~\cite{oppenheim_uncertainty_2010} and that numerical 
data show that $I$ monotonically decreases with $d$.

\section{\label{sec:Stability} Robustness of Generalized Bell Inequalities}

\begin{figure}
\includegraphics[width=0.44\textwidth]{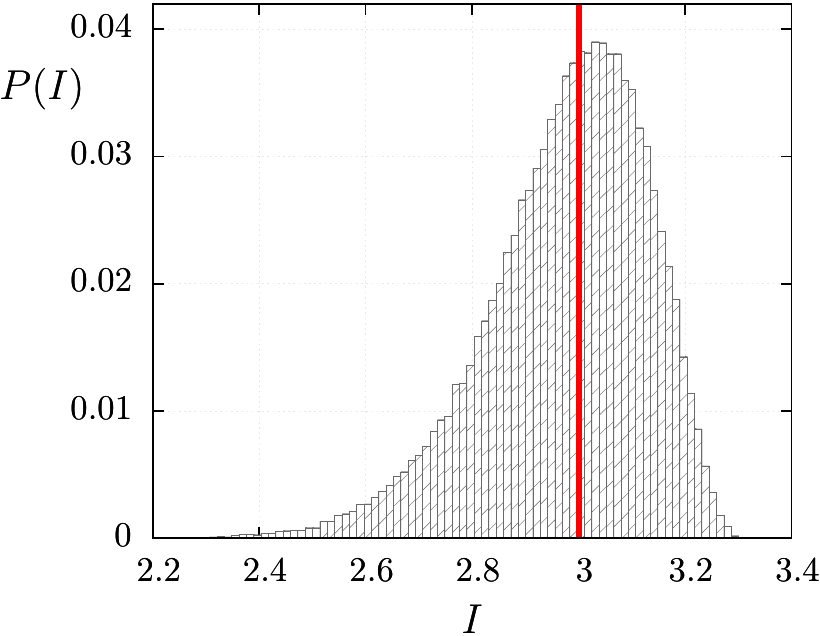}
\caption{(Color online) \label{fig3} Histogram of the probabilities $P(I)$ to obtain the values $I$ for the maximally entangled state given in Eq.~\eqref{bstate} with 
$l=1$ for a perturbation $\varepsilon \approx 0.233$. The red line $I=3$ indicates the border between classical ($I\le3$) and non-classical correlations ($I > 3$).}
\end{figure}

In this Section we characterize as a function of the system size 
the robustness of the above generalized Bell inequalities 
against perturbations of the projections identified in Eqs.~(\ref{eq:optAxes}). 
This is indeed a fundamental question to address when investigating the possibility to violate a Bell-like inequality using a macroscopic quantum system. In fact, any experimental test of this kind will have to cope with non-idealities in the identification of the most suitable projections to implement. Our study is thus aimed at determining the tolerance allowed by the Bell-like tests addressed above against imperfect measurement settings. 

\begin{figure}
\includegraphics[width=0.44\textwidth]{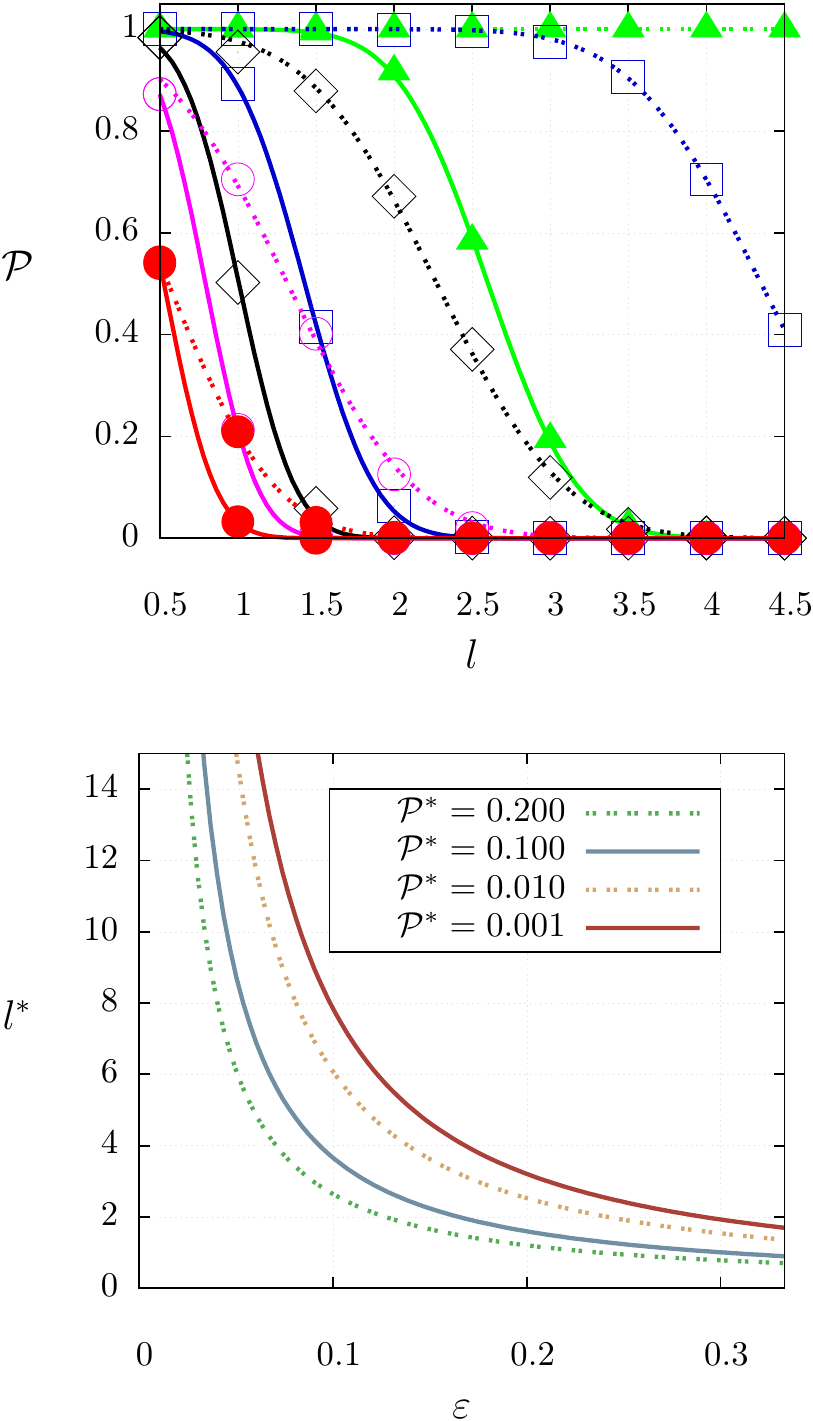}
\caption{(Color online) \label{fig2} Upper panel: 
Violation probability $\mathcal{P}_B$ as a function of 
the spin quantum number $l$ for  $\varepsilon= 0.12,0.18,0.23, 0.29,0.41$ (green triangles, blue squares, black diamonds, magenta circles, red filled circles). 
Solid and dashed curves show the fit (\ref{eq:errFunctFit}) for $I$ and $I_d$, respectively.
Lower panel: estimate from Eq.~(\ref{eq:fit_Lstar_epsilon}) of the threshold 
$l^*_I$ as a function of $\varepsilon$, for different threshold probabilities $\mathcal{P}_I^*$ (see legend).}
\end{figure}

In order to attack this problem, we consider the equivalence between an improperly arranged measurement setting [i.e., projections of the ideal generalized Bell state performed on states that differ from those in Eqs.~(\ref{eq:optAxes}), but still being mutually orthogonal] and the performance of ideal measurements on a state that is locally rotated by an unknown amount and around an unspecified direction. We thus perform a stability analysis in the regime of small perturbations by independently rotating each subsystem by an angle $\varepsilon$ around a random axis. The rotations are performed by applying the bi-local unitary transformation 
\begin{equation}
\label{unitOp}
 \hat U= \hat U_A \otimes \hat U_B = e^{\mathrm{i} \varepsilon  \hat H_A}  \otimes e^ {\mathrm{i} \varepsilon  \hat H_B},
\end{equation}
where $\hat H_\mu$  ($\mu=A,B$) is a $d$-dimensional Hermitian operator whose entries $(H_{\mu})_{ij}$ are randomly chosen from a uniform distribution of values within the range $-1\le|(H_{\mu})_{ij}|\le1$.

We have performed an extensive numerical analysis of the value of the Bell expressions $I$ and $I_d$ within the framework discussed above.
In order to obtain the probability distributions $P(B)$ for $B=I, I_d$ 
we have accumulated statistics by means of $N=10^5$ random rotations for each value of $\varepsilon$ and $d$ considered in our study. 
A typical example of the results of this analysis is reported, for $B=I$, 
in Fig.~\ref{fig3}:  the distribution $P(I)$ is a Gaussian-like distribution with a negative skew. 
In order to provide a quantitative characterization of the robustness
of the Bell inequalities, we focus on the violation probability 
\begin{equation}
\mathcal{P}_{B}= \int_{B_c}^\infty P(B') dB',
\label{eq:PBstar}
\end{equation}
where $B_c= 2,3$ for $B,B'=I_d,I$, respectively. Such quantity is 
the area of the distribution $P(B)$ at the rightmost 
side of the LHV value $B_c$. 

We have studied the behavior of the the probability 
$\mathcal{P}_B$ 
against the dimension of the problem at hand: 
in Fig.~\ref{fig2} (top panel) we plot $\mathcal{P}_B$ as a function of 
the spin quantum number $l$ ($d=2l+1$), for various choices of 
$\varepsilon$ and for both $I$ and $I_d$.
We first note that for $l=1/2$ the results for the two 
Bell expressions coincide. 
On the contrary, from $l =1$ onwards, 
$\mathcal{P}_{I_d}> \mathcal{P}_{I}$,
namely violation of the Bell inequality for $I_d$, which includes number of terms growing with the dimension of the system, is more resistant to local perturbations than violation for $I$. 
It is clear from the top panel of Fig.~\ref{fig2} that for all values of
$\varepsilon$ that we have considered and for a given 
minimum violation threshold $\mathcal{P}_B^*$, one can find a 
threshold value $l^*_B(\varepsilon,\mathcal{P}_B^*)$ such that 
$\mathcal{P}_B<\mathcal{P}_B^*$ for all 
$l>l^*_B$. In order to quantify $l^*_B$, we fit
the data of Fig.~\ref{fig2} (top panel) with the function 
\begin{equation}
  \label{eq:errFunctFit}
  \frac{1}{2} \left[ 1 - \erf\left(\frac{l - \bar{l}_B(\varepsilon)}{\Delta_B(\varepsilon)} \right) \right]\text{,}
\end{equation}
characterized by the mean value $\bar{l}_B(\varepsilon)$ and the standard deviation $\Delta_B(\varepsilon)$. 
This allows us to estimate  $l^*_B$ as
\begin{equation}
\label{eq:fit_Lstar_epsilon}
 l^*_B(\varepsilon,\mathcal{P}_B^*) \approx 
\bar{l}_B(\varepsilon)
+ \Delta_B(\varepsilon) \erf^{-1} (1 - 2 \mathcal{P}^*_B), 
\end{equation}
with $\bar{l}_B(\varepsilon)\approx 0.17\varepsilon ^{-0.61}$ and
$ \Delta_B(\varepsilon)\approx 0.137\varepsilon^{-1.38}$ for $B=I$;
$\bar{l}_B(\varepsilon)\approx 0.057\varepsilon^{-2.52}$ and
$ \Delta_B(\varepsilon)\approx 0.484\varepsilon^{-0.50}$ for $B=I_d$.
In Fig.~\ref{fig2} (bottom panel) we report 
$l^*_I$ for several values of $\mathcal{P}^*_I$.
These results clearly indicate that even for
small error $\varepsilon$, the violation probability becomes
rapidly negligible when increasing $l$.

\begin{figure}[t]
\includegraphics[width=0.44\textwidth]{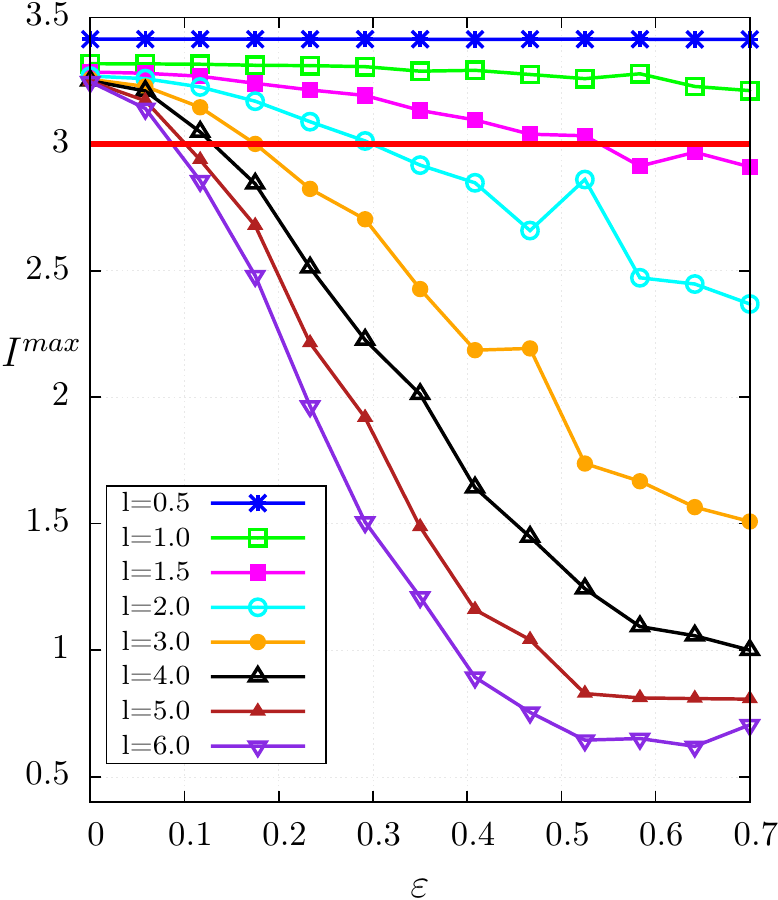}
\caption{(Color online) \label{fig4} Maximal value $I^{\rm max}(\varepsilon)$ for different spin quantum number $l$. Red horizontal line indicates the violation bound $I = 3$.}
\end{figure}

Besides studying the probability to violate a given inequality, it is interesting to consider 
whether or not one is actually able to violate a Bell inequality
in a $d$-dimensional system, 
for a given $\varepsilon$ and a finite sampling.
This is equivalent to estimate the quality of the statistical sampling of the tails of the Gaussian-like distribution $P(B)$. In order to do so, we adopt a pragmatic approach according to which we consider it possible to violate inequality for the Bell expression $B$ if, with a sampling of $N=10^5$ events, a single instance of violation is found. We then extract the maximum occurred value $B^{\rm max}$ of the Bell expression and study it against $\varepsilon$ for increasing system dimensions. Fig.~\ref{fig4} shows the results for the case of $B=I$, highlighting the existence of a value $\varepsilon^*_I$ above which the corresponding inequality is no longer violated. Although the quantitative analysis is not reported here, we have found that an analogous critical value  $\varepsilon^*_{I_d}$ exists for the case of $B=I_d$, too.

\begin{figure}[t]
\includegraphics[width=0.44\textwidth]{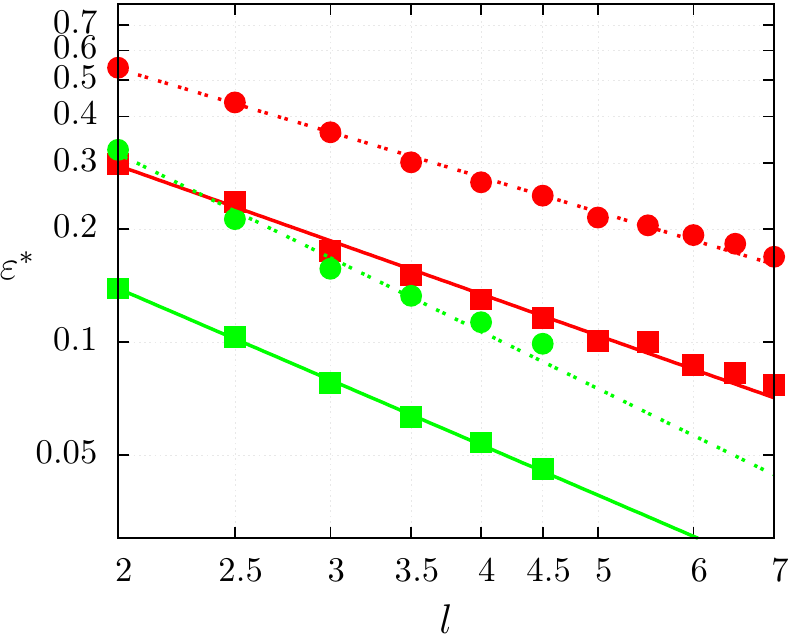}
\caption{\label{fig5} Log-log plot of critical $\varepsilon_B^*(l)$ with corresponding fits (form $a/l^b$) for $B=I$ (solid red with squares) and $B=I_d$ (dashed red with circles). The solid green line with squares denotes the global critical $\varepsilon_{I,g}^*$ for global rotations
for inequality $I$, the dashed green line with circles is $\varepsilon_{I_d,g}^*$ for inequality $I_d$.}
\end{figure}

The values of both $\varepsilon^*_I$ and $\varepsilon^*_{I_d}$ at given $l$ are reported in Fig.~\ref{fig5}, which also displays the comparison with the fitting function $a/l^b$, thus showing a power-law  decay of the error threshold with the size of the system, 
with $b\approx -1.13$ for $I$ and $b\approx -0.96$ for $I_d$.
Consistently with our findings on the probability to violate a given inequality, we find that $I_d$ is more robust than $I$ as the corresponding critical value $\varepsilon^*_{I_d}$ decays with $l$ more slowly than $\varepsilon^*_I$. 

We complete this Section by studying the stability of the generalized Bell inequalities when global perturbations are used instead of the bi-local ones considered so far. By repeating the same analysis illustrated above under the assumption that the random perturbation now acts on both the spins of our system, thus possibly deteriorating their shared correlations, we find the results
reported by the green lines in Fig.~\ref{fig5}. A global rotation of the state of the system results in a more pronounced decay of the threshold value of $\varepsilon$ that, however, does not affect significantly the scaling with the size of the system. 
From the same power law fit $a/l^b$ as above we obtain 
$b\approx -1.39$ for $I$ and $b\approx -1.57$ for $I_d$. 
For random global rotations, it can be shown analytically 
(see appendix \ref{app:levy}) that the 
asymptotic (in $d$) decay of of the violation probability 
$\mathcal{P}_{I_d}$ is at least exponentially fast. 

\section{\label{sec:Optimization} Random measurement operators}

In this section, we consider the violation of generalized Bell inequalities by a bipartite spin-$d$ system  prepared in an unknown state. In order to make our study more specific and concrete, we first consider a spin system prepared in a maximally entangled state but defined over random $z$-axes. 
In such a case, it is no longer possible to identify the form of the optimal measurement operators $\hat A$s and $\hat B$s, and the measurement settings given in Eqs.~\eqref{eq:optAxes} would be unsuitable for the successful falsification of a given generalized Bell inequality. 
In order to address this point quantitatively, we resort to the observation that the picture given above is fully equivalent to that of a spin system prepared in the ideal generalized Bell state and subjected to unknown local measurement settings. 

We thus perform measurements along random directions and check if this procedure allows to detect a violation with a statistically relevant probability. This procedure is not equivalent to the analysis we have done in the previous Section, as the {\it direction} of the projections entailed by the measurement steps needed for the construction of the Bell function are now fully random and not mutually orthogonal. Every measurement operator is rotated individually.

\begin{figure}[t]
\includegraphics[width=0.44\textwidth]{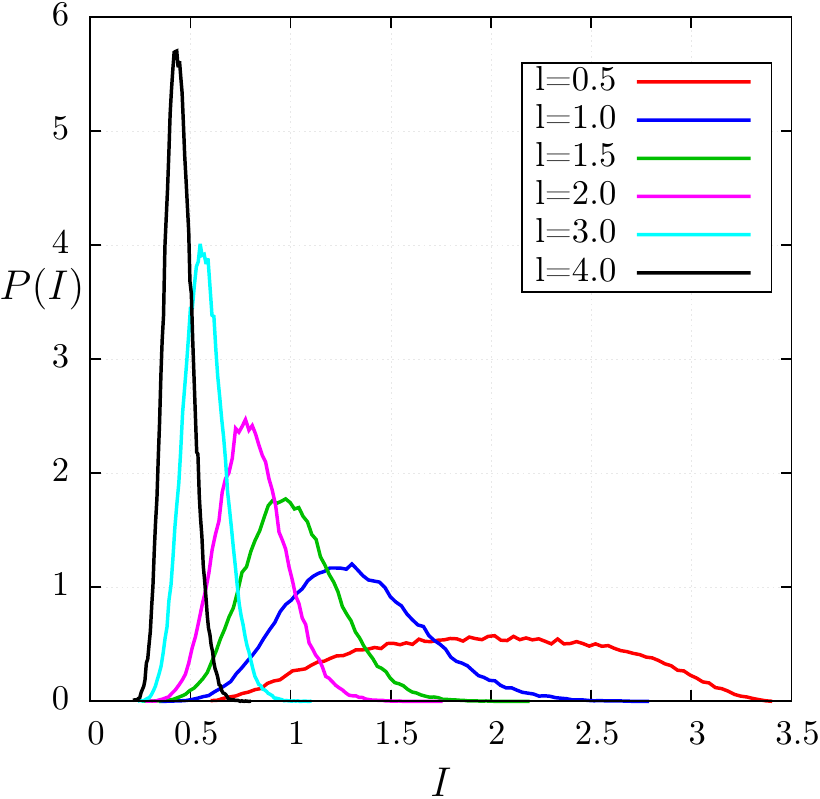}
\caption{(Color online) \label{fig6} Histograms $P(I)$ of the Bell inequality $I$ for Bell state $|\Phi^+\rangle$ and random measurement operators ${A}_1$, ${A}_2$, ${B}_1$, and ${B}_2$, for various
spin numbers $l$.}
\end{figure}

In Fig.~\ref{fig6} we report the resulting distributions of the values taken by $I$ for various dimensions of the system. The distributions can be well fitted by Gaussian functions specified by 
mean value $\bar{I}$ and standard deviation $\sigma$ decreasing with $l$. Specifically, we find the best fit values $\bar{I} \propto l^{-0.79}$ and $\sigma \propto l^{-1.13}$. This implies that the probability of measuring a violation 
is as small as $\mathcal{P} \approx 4 \cdot 10^{-7}$ for $l=1$,
and is below machine precision for any $l>1$. This is an interesting result, as it entails that, for systems with four or more local states, it is virtually impossible to violate a generalized Bell inequality, even when operating over a maximally entangled state but with random direction.

\begin{figure}[t]
\includegraphics[width=0.44\textwidth]{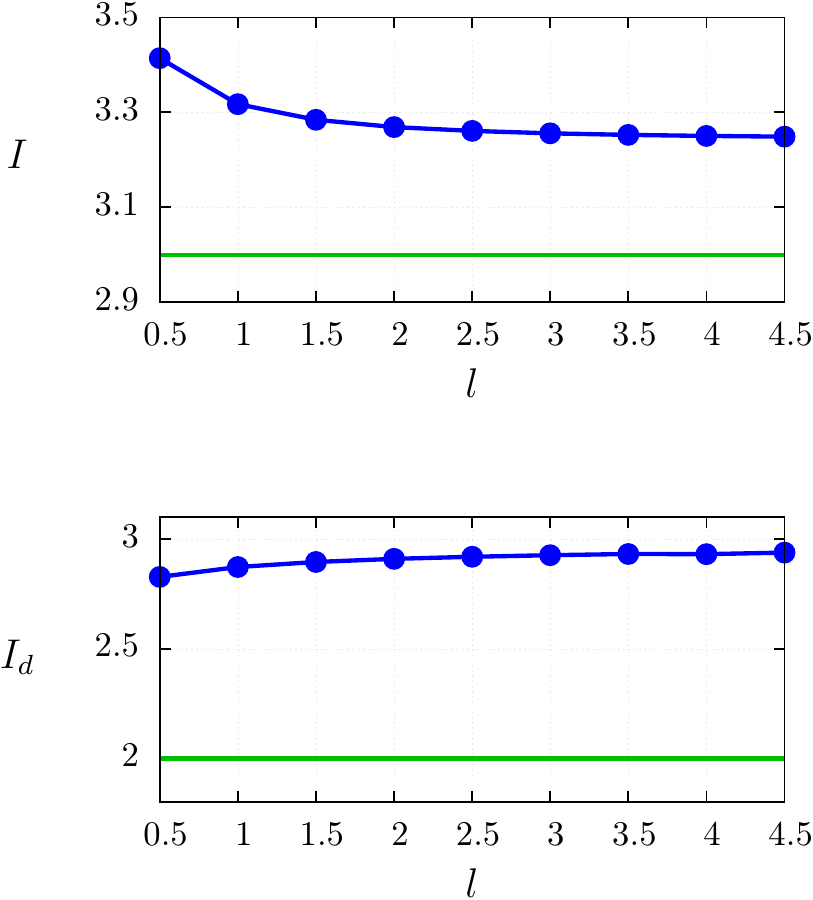}
\caption{(Color online) \label{fig7} Optimized values $I^{\rm opt}$ (upper panel) and $I_d^{\rm opt}$ (lower panel) for different spin quantum numbers $l$ (blue curve) and maximally entangled state $|\Phi^+\rangle$. 
In both panels the  bottom (green) line corresponds 
to the border between classical and non-classical correlations.}
\end{figure}

Given that random search fails to detect violation in general, we adopt a more sophisticated strategy and optimize the observables to be measured, employing the Nelder-Mead Simplex  algorithm~\cite{nelder_simplex_1965}, which is a general-purpose gradient-free algorithm of vast use in complex multi-variable searches. We repeat the optimization from different  randomly chosen initial sets of coefficients for the matrix representation of the observables $A$s and $B$s and report in Fig.~\ref{fig7} the maximum achieved 
value of $I$ and $I_d$.
As it can be seen, in both cases a violation of the Bell inequalities is found for systems of up to $l=4.5$ and there is indication (as it should be) that for all system sizes it should be possible to find a set of observables capable to detect the violation. However, the computational time of one function evaluation scales as $\mathcal{O}(l^6)$ for inequality $I$ and as $\mathcal{O}(l^7)$ for $I_d$ \cite{footnote1}.
In combination with an increasing number of function evaluations needed for increasing $l$, this once more practically 
prevents the effectiveness of such protocol for increasing system sizes
\cite{footnote}. 

\begin{figure}[t]
\includegraphics[width=0.4\textwidth]{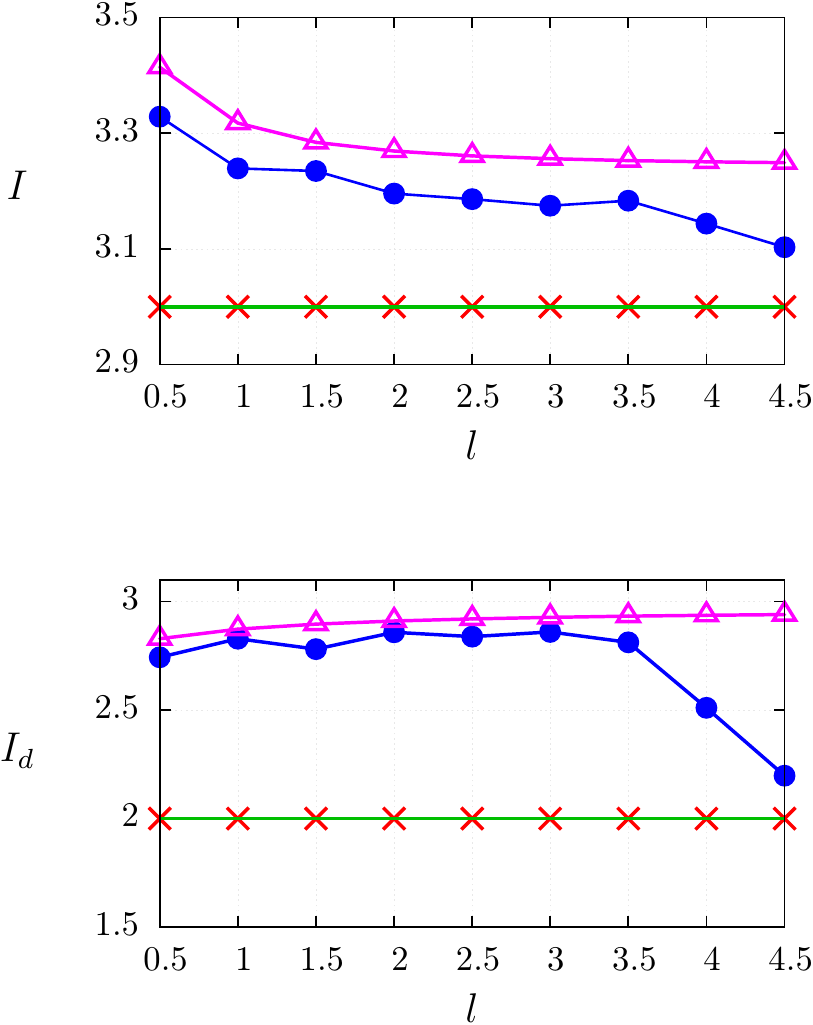}
\caption{(Color online) \label{fig8}  Optimal values for Bell expressions $I$ 
(upper panel) and $I_d$ (lower panel), for random entangled states (blue curve) 
and random separable states (crosses).   
Green line: bound for classical correlations,
Triangles: values for $I$ and $I_d$ for the maximally entangled state  
and optimal measurement setting as reported in Fig.~\ref{fig1}. Note: Blue points are always the maximal value of the five random states as described above.}
\end{figure}

As a final step in our study, we now check how the violation of the generalized Bell inequalities behaves for 
random two-spin states. We considered two cases:
\begin{enumerate}
 \item Random entangled states whose vectors have elements randomly chosen from a uniform distribution of values within the range $[-10,10]$ and then properly normalized.
 \item Product states 
 of two random $d$-dimensional states 
 (built up as described above). 
\end{enumerate}
For each value of $l$ we consider five different random entangled states and two random product states and store the maximally violating instance for both cases.  
Typical results for maximally violating random entangled and a random product state are shown in Fig.~\ref{fig8}, where we observe that violations are possible for both $I$ and $I_d$ and arbitrary values of $l$ (blue circles). 
This of course is only true for random entangled states, as for random product states (red crosses) a violation is impossible. Nevertheless, the critical bounds can always be reached by classical random states.
Note that once again, finding the proper setting to violate the Bell inequalities get harder for bigger system sizes: the cases for $l \geq 3.5$ in Fig.~\ref{fig8} display a lower violation with respect to the other ones. 
However, this is due only to the fact that we stop the optimization after one week of computational time, which in these cases is clearly not enough to achieve the maximal possible violation, demonstrating 
once more the increasing difficulties in finding a violation for increasing system size.  

\section{Conclusions}
\label{conc}

We have analyzed quantitatively the violation of Bell inequalities formulated for general $d$-dimensional systems. Our approach, which is oriented towards the provision of information relevant to experimentalists, was based on the careful evaluation of the impact that imperfections in both the state resource and the measurement settings could have on actually violating one of such generalized non-locality tests. We have found that both bi-local and global imperfections in the arrangements for the measurement settings needed to test a given inequality affect the quality of the corresponding Bell test in a system size-dependent fashion. Non-locality in bipartite systems spanning large Hilbert spaces is difficult to infer, even when managing maximally entangled states of two spin-$l$ particles. Interestingly, we have identified size-dependent thresholds on the amount of noise that the system is able to tolerate before rendering the chance of violating a generalized Bell inequality negligibly small. We have also 
highlighted 
the influences due to the statistical sampling used to check the falsification of the Bell tests addressed in our work: while a uniform random sampling of the measurement sampling makes it unlikely to single out an instance of violation even for qutrit systems, a more sophisticated statistical sampling reveals that it should be possible to find such instances regardless of the size of the local subsystems, although with increasing efforts being required. 

We believe that this work helps addressing the inherent complexity of violating Bell inequalities above and beyond the usual studies used to assess their robustness, and provides useful information for any experimental endeavor aimed at testing non-local features in the state of systems spanning large Hilbert spaces, a situation that is quickly becoming more realistic and addressable. Furthermore, our work provides for a transparent interpretation of the emergence of classicality in terms of local realism at the macroscopic level: without resorting to decoherence mechanisms, the size-dependent noise thresholds we identified above describe the practical impossibility to detect non-classical behavior in large enough systems via Bell inequalities violations. In a word, in the unavoidable presence of noise, large systems will behave for all practical purposes classically under a Bell test.

Numerous avenues are left open by the questions addressed in this work, the most natural being whether the use of non-maximally entangled states leads to a less demanding inference of non-locality in large-sized 
systems~\cite{znidaric}. In this context, it would be interesting to explore 
the robustness of the so-called {\it embezzling states}~\cite{embe'}, 
i.e. of a family of partially entangled states which exhibit 
maximal violation of Bell inequalities, under the random errors 
addressed herein.

\section{Acknowledgements}
\label{ackl}

We thank P. Silvi and M. Keck for discussions. We acknowledge support from the EU via the SIQS and RYSQ project; the DFG via the SFB/TRR21, the BMBF (QUOREP) and the MIUR-PRIN; and the BWgrid for computational resources. MP is supported by the EU FP7 grant TherMiQ (Grant Agreement 618074), the John Templeton
Foundation (Grant ID 43467), and the UK EPSRC (EP/M003019/1).  

\appendix
\section{Asymptotic decay of $\mathcal{P}_{I_d}$}

\label{app:levy}

$I_d$, as defined in Eq.~(\ref{Idbell}), is a function of the (pure) state $\psi$ of the bipartite system, hence a function on the $2d^2-1$ dimensional real unit sphere $S_{2d^2-1}$. It will be shown that this function is subject to the ``concentration of measure'' phenomenon \cite{Levy}, meaning that, if $\psi$ is chosen at random from the uniform distribution  on $S_{2d^2-1}$, then the probability that $|I_d(\psi)|$ is larger than an arbitrarily small positive number is exponentially small with $d$; notably,
\begin{equation}
\label{main}
 \mathcal{P} \left\{|I_d(\psi)|\geq\epsilon\right\}\;\leq\;2\exp\left(-\frac{d^2\epsilon^2}{192(6+d)\pi^3}\right)\;.
\end{equation}

To show this it is convenient to rewrite $I_d$  in the form:
\begin{gather}
\label{iddi}
I_d(\psi)\;=\;R_d(\psi)\;-\;S_d(\psi)\;,\nonumber\\
R_d(\psi)\;=\;\sum\limits_{1\leq i,j\leq 2}\sum\limits_{k=0}^{s-1}c(k)\;\|{\hat P}^{(i,j)}_k\psi\|^2
\;,\nonumber\\
S_d(\psi)\;=\;\sum\limits_{1\leq i,j\leq 2}\sum\limits_{k=0}^{s-1}c(k)\;\|{\hat P}^{(i,j)}_{-k-1}\psi\|^2\;,
\end{gather}
where $s=[d/2]$ , $c(k)=1-2k/(d-1)$. For $i,j=1,2$ and  $-d+1\leq k\leq d-1$,
$
{\hat P}^{(i,j)}_k$  is the projector onto the d-dimensional subspace wherein $A_i-B_j=n(i,j,k)$,
where :
$$
n(1,1,k)=k\;,\;\;n(1,2,k)=-k\;,
$$
$$
n(2,1,k)=-k-1\;,\;\;n(2,2,k)=k\;.
$$
The average of $I_d(\psi)$ with respect to the uniform, normalized measure on $S_{2d^2-1}$ is $0$,
 because projectors which share the same superscript $i,j$  in $R_d$ and in $S_d$   are unitarily equivalent and the uniform measure is unitarily invariant. Then Levy's lemma states that \cite{Levy}
 \begin{equation}
 \label{levy}
 \mathcal{P} \left\{|I_d(\psi)|\geq \epsilon\;\right\}\;\leq\;2 \exp\left(-\frac{d^2\epsilon^2}{9\pi^3\ell_d^2}\right)\;,
 \end{equation}
 where $\ell_d$ is the Lipschitz constant of $I_d$, to be presently estimated. First we write:
 \begin{gather}
 |I_d(\psi)\;-\;I_d(\phi)|\;\leq\;|R_d(\psi)-R_d(\phi)|\;+\;|S_d(\psi)-S_d(\phi)|\;,
 \end{gather}
 and we note that, for an arbitrary projector $\hat P$ and arbitrary normalized states $\psi,\phi$
 $$
 |\;\|\hat P\psi\|^2\;-\;\|\hat P\phi\|^2\;|\;\leq\;2\|\hat P(\psi-\phi)\|\;.
 $$
 It follows that:
 $$
 |I_d(\psi)-I_d(\phi)|\;
 $$
 \begin{equation}
 \label{here}
\leq\;2\sum\limits_{1\leq i,j\leq 2}\sum\limits_{k=0}^{s-1}c(k)\;(\|\hat P_k^{(i,j)}(\psi-\phi)\|\;+\;\|\hat P^{(i,j)}_{-k-1}(\psi-\phi)\|).
 \end{equation}
 In the above sum, projectors $\hat P^{(i,j)}_k$ that share the superscript $(i,j)$ with different $k$ are mutually orthogonal, so
 $\hat P_k^{(i,j)}+\hat P_{-k-1}^{(i,j)}$ is again a projector $\hat R^{(i,j)}_k$, and
 $$
 \|\hat P^{(i,j)}_k(\psi-\phi)\|+\|\hat P^{(i,j)}_{-k-1}(\psi-\phi)\|\;\leq\;\sqrt{2}\|\hat R^{(i,j)}_k(\psi-\phi)\|\;;
 $$
 using this in (\ref{here}) we get
 \begin{gather}\
 |I_d(\psi)-I_d(\phi)|\;\leq\;2\sqrt{2}\sum\limits_{1\leq i,j\leq 2}\sum\limits_{k=0}^{s-1}c(k)
 \;\|\hat R^{(i,j)}_k(\psi-\phi)\| \nonumber\\
 \leq 2\sqrt{2}\sum\limits_{1\leq i,j\leq 2}\left(\sum\limits_{k=0}^{s-1}c(k)^2\right)^{1/2}\left(
 \sum\limits_{k=0}^{s-1}\|\hat R^{(i,j)}_k(\psi-\phi)\|^2\right)^{1/2}\nonumber\\
 \leq 8\sqrt{2}(1+s/3)^{1/2}\|\psi-\phi\|\;.
 \end{gather}
In the last line, we used that $\hat R^{(i,j)}_k$ with the same $i,j$ and 
different $k$ are orthogonal.
This inequality shows that $\ell_d\leq 8\sqrt{2}(1+s/3)^{1/2}$. Using this in Levy's lemma
 yields the announced estimate.

 A sharper bound is obtained by using the following version of Levy's lemma:
 \begin{equation}
 \label{levy2}
 \mathcal{P} \{|I_d-M_d|>\epsilon\}\;\leq\;\exp(-d^2\epsilon^2/(2\ell^2_d))\;,
 \end{equation}
 where the median  $M_d$ is defined such that Prob$\{I_d\leq M_d\}=1/2$. While we have no theoretical estimate for $M_d$, numerical results suggest that, at least with the observables we have chosen, the distribution of $I_d$ is even. This  would imply $M_d=0$, and then:
 $$
  \mathcal{P} \{|I_d|>\epsilon\}\;\leq \;\exp(-3d^2\epsilon^2/(128(6+d)))\;.
 $$

While even this stricter bound 
becomes relevant only for dimensions 
$d\gtrsim 10$, bound (\ref{main}) proves
rigorously that the violation probability
$\mathcal{P}_{I_d}$, upper bounded by the right-hand side of 
(\ref{main}) for $\epsilon=2$, decays at least exponentially fast with $d$.


\begin{thebibliography}{99}
\bibitem{footnote1}
The longest computational time when computing one probability $P$ is needed for a matrix-matrix product, scaling with $\mathcal{O}(N^3)$, 
where $N \sim d^2$. For inequality $I_d$ not only $4$, but $\mathcal{O}(d)$ probabilities have to be calculated, resulting in a $\mathcal{O}(l^7)$ behavior.
\bibitem{footnote}
The results presented here have been computed with optimisations lasting a few days for each set of system parameters, on a 
state-of-the art computational cluster.
\end{thebibliography}
\end{document}